\documentclass[twocolumn,showpacs,showkeys,preprintnumbers,amsmath,amssymb,prd]{revtex4}

\usepackage{graphicx}
\usepackage{dcolumn}
\usepackage{bm}
\begin{document}


\title{Nonassociative geometry: Friedmann-Robertson-Walker spacetime}

\author{Alexander I. Nesterov}
\affiliation{Departament of Physics, C.U.C.E.I., Guadalajara University,
Guadalajara, Jalisco, Mexico}
\email{nesterov@cencar.udg.mx}

\author{ \boxed{\rm \;Lev \; \,V.\;\, Sabinin}}
\affiliation{Faculty of Science, UAEM, Cuernavaca, Morelos, Mexico}
\email{lev@servm.fc.uaem.mx}


\begin{abstract}
In (Phys. Rev. {\bf D 62}, 081501(R), (2000)) we proposed  a unified approach
to description of continuous  and discrete spacetime based on nonassociative
geometry and described nonassociative smooth and discrete de Sitter models. In
our paper we give the description of nonassociative  Friedmann-Robertson-Walker
spacetime.

\end{abstract}

\pacs{PACS numbers: 04.20.-q; 04.20.Gz; 02.40.Hw } \keywords{nonassociative
geometry, discrete spacetime, loop, quasigroup}

\maketitle

Recently \cite{NS1,NS} we have proposed  in the framework of {\em
nonassociative geometry} \cite{S5,S6,S7,S8,NS} a new approach to the
description of continuous and discrete spacetime. It follows from our our
approach that at distances comparables with Planck length the standard concept
of spacetime might be replaced by the nonassociative discrete structure, and
nonassociativity is the algebraic equivalent of the curvature.

In \cite{NS1} the general formalism  has been applied to the case of de Sitter
space. In this paper we study the nonassociative continuous and discrete
Friedmann-Robertson-Walker (FRW) models.
~~~~~\\

\noindent {\bf Mathematical preliminaries}. The foundations of nonassociative
geometry are based on the fact that in a neighborhood of an arbitrary point on
a manifold with an affine connection one can introduce the geodesic local loop,
which is uniquely defined by means of the parallel translation of geodesics
along geodesics. Here the main emerging algebraic structures are related to the
theory of quasigroups and loops (for details see: \cite{K,S1,S2,S8,NS,NS1}).

{\em Definition 1.} Let $\langle Q,{\mathbf\cdot}\rangle $ be a groupoid with a
binary operation $(a,b) \mapsto a {\mathbf\cdot} b$ and $Q$ be a smooth
manifold. Then $\langle Q,{\mathbf\cdot}\rangle $ is called a {\it quasigroup}
if each of the  equations $a{\mathbf\cdot} x=b,~y{\mathbf\cdot} a=b$ has a
unique solution:  $x=a\backslash b$, $y=b/a$. A {\it loop} is a quasigroup with
a two-sided identity, $a{\mathbf\cdot} e= e{\mathbf\cdot} a=a, \forall a \in
Q$. A loop $\langle Q,{\mathbf\cdot},e \rangle$ with a smooth functions
$\phi(a,b):=a{\mathbf\cdot} b$ is called a {\it smooth loop}.

Nonassociativity of the operation in a loop  $\langle
Q,{\mathbf\cdot},e\rangle$ is described by an {\em associator} $l_{(a,b)}$ such
that the following identity holds
\begin{equation}\label{L1a}
a{\mathbf\cdot} (b{\mathbf\cdot} c) =(a{\mathbf\cdot} b)l_{(a,b)}c.
\end{equation}
Introducing a left translation $L_a$ as the following: $L_a b =a{\mathbf\cdot}
b$, we obtain
\begin{equation}
l_{(a,b)}=L^{-1}_{a{\mathbf\cdot} b}\circ L_a\circ L_b.
\label{Ll}
\end{equation}

{\em Definition 2.} Let $\langle{ M},{\mathbf\cdot},e \rangle$ be a
partial {\it groupoid} with a binary operation $(x,y)\mapsto
x{\mathbf\cdot} y$ and a neutral element $e ,\; x{\mathbf\cdot} e  =e
{\mathbf\cdot} x =x$; $   M$ be a smooth manifold (at least $C^1$-smooth)
and the operation of multiplication (at least $C^1$-smooth) be defined in
some neighborhood $U_e $, then $\langle { M},{\mathbf\cdot},e \rangle$
is called a {\it partial loop} on $ M$.

{\em Definition 3.} Let $\langle{ M},{\mathbf\cdot},e \rangle$ be a left
loop with a neutral element $e$ and $t: x \mapsto tx$ be a unary operation
such that $(t+u)x = tx{\mathbf\cdot} ux$, $(tu)x = t(ux)$ $(t,u \in
{\mathbb R}, \; x \in M)$. Then $\langle M,{\mathbf\cdot},e,(t)_{t\in
\mathbb R}\rangle $ is called a {\em left odule}.

{\em Definition 4.} Let $M$ be a smooth manifold with a smooth partial ternary
operation, $x\;{}_{\stackrel{\mathbf\cdot}{a}}\;y= L^a_x y\in { M}$ such that
$L^a_x y$ defines  in the some neighbourhood of the point $a$   the loop with
the neutral $a$, then the family $\bigl \{\langle
M,\;{}_{\stackrel{\mathbf\cdot}{a}}\; \rangle\bigr\}_{a\in M}$ is called a {\it
loopuscular structure}.

Let $M$ be a $C^k$-smooth $(k\geq 3)$ affinely connected manifold and the
following operations are given  on $M$:
\begin{align}
& L^a_x y= {x\; {}_{\stackrel{\cdot}{a}}\; y}
={\rm Exp}_x \tau^a_x {\rm Exp}^{-1}_a y, \\
& \omega_t(a,z) = t_a z= {\rm Exp}_a t {\rm Exp}^{-1}_a z, \\
&{x\; {}_{\stackrel{+}{a^{}}}\; y}
={\rm Exp}_a ({\rm Exp}^{-1}_a x + {\rm Exp}^{-1}_a y),
\end{align}
${\rm Exp}_x$ being the exponential map at the point $x$ and $\tau^a_x$ the
parallel translation along the geodesic going from $a$ to $x$. This
construction equips $M$ with the so-called {\em natural linear geodiodular
structure of an affinely connected manifold} $({M},\nabla)$.

If $\bigl \{\langle  M,\;{}_{\stackrel{\mathbf\cdot}{a}}\;
\rangle\bigr\}_{a\in M}$ is a loopuscular structure then
\begin{equation}
h^a_{(b,c)} = (L^a_c)^{-1}\circ L^b_c \circ L^a_b
\label{hol}
\end{equation}
is called the {\em elementary holonomy}. It satisfies the odular Bianchi identities:
\begin{equation}
h^a_{(z,x)}\circ h^a_{(y,z)}\circ h^a_{(x,y)} = (L^a_x)^{-1}\circ
h^x_{(y,z)} \circ L^a_x
\label{B1},
\end{equation}
and in the linear approximation reduces to the conventional Bianchi identities.

The family of local geodesic loops considered above uniquely defines the space
with affine connection and  there exist some relations between the loops at
distinct points that can be expressed by means of some algebraic identities,
the so-called {\em geoodular identities}.

Actually, any $C^k$-smooth $(k\geq 3)$  affinely connected manifold can be
considered as a geodiodular structure and the categories of geoodular
(geodiodular) structures and of affine connections are equivalent
\cite{S8,S2,S3,S4}.
\\

{\em Nonassociative space}. -  Nonassociative geometry is based on the
construction considered above. A geoodular covering of the affinely connected
manifolds, consisiting of the odular covering with the geoodular identities,
contains complete information about the manifold and allows us to reconstruct
it. In fact, if we take an arbitrary smooth geoodular covering, then it
uniquely generates an affine connection, whose geoodular covering coincides
with the initial one. Ignoring the smoothness, it is possible to consider
discrete spaces, introduce ``metric'' and ``connnection'' over arbitrary fields
or rings, and even to define finite spaces with ``affine connection''.

We say that ${\mathcal {M}}= \bigl\{\langle  M,\; {}_{\stackrel{\mathbf\cdot
}{a}}\;,{}_{\stackrel{\mathbf +}{a}}\;,(t_a)_{t\in {\mathbb R}}
\;\rangle\bigr\}_{a\in M}$ is a {\em nonassociative space}, if there exists
geodiodular covering of $M$ such that:

1. $\langle M,\;{}_{\stackrel{\mathbf \cdot}{a}}\;,(t_a)_{t\in \mathbb
R}\rangle $ is an odule with a neutral $a\in M$, for any $a\in M$,

2. $\langle M,\;{}_{\stackrel{\mathbf +}{a}}\;,(t_a)_{t\in \mathbb R}\rangle $
is a $n$-dimensional vector space (with zero element $a\in M)$,

3. The following geoodular identities hold
\begin{align}
&L_{u_ax}^{t_ax}\circ L_{t_ax}^{a}=L_{u_ax}^{a},\nonumber \\
&L_{x}^{a}\circ  {t_a}=t_x \circ  L_{x}^{a},\nonumber \\
&L_{x}^{a}(y\;{}_{\stackrel{\mathbf +}{a}}\; z)= L_{x}^{a}y \;
{}_{\stackrel{\mathbf +}{x}}\; \;L_{x}^{a}z  \nonumber
\label{gid3}
\end{align}
One may consider the operations above as partially defined.

In our approach the so-called {\em osculating space} ${\mathcal
{M}}^+_a=\langle M,{}_{\stackrel{\mathbf +}{a}}, a ,(t_a)_{t\in \mathbb
R}\rangle$ plays the role of tangent space at $a\in M$. The presence of
curvature in a nonassociative space results in a non-trivial elementary
holonomy,
\[
L_a^y\circ  L_y^x\circ  L_x^a= h^a(x,y)\neq \rm {Id}.
\]
In addition, one can enrich $\mathcal M$  by a {\em metric diodular structure},
introducing left invarint metric $g^a (x,y)$ as follows:
\begin{equation}\label{G1}
    g^b (L_b^a x,L_b^a y)=g^a(x,y).
\end{equation}

~~~\\

\noindent {\bf Nonassociative FRW spacetime}. In what follows  we will give the
description of the continuum and discrete FRW spacetime in the framework of
nonassociative geometry.

The metric of the general Friedmann-Robertson-Walker (FRW)model may be written
as \cite{HE,KR}
\begin{equation}\label{g1a}
ds^2  = d\tau^2 - a^2(\tau)\Big(dr^2 + \Sigma^2(r,K)\big(d\theta^2 +
\sin^2\theta d\varphi^2\big)\Big),
\end{equation}
where $\Sigma(r,K)= \sin r,\; r$, or $\sinh r$, respectively when $K=1,0,$ or $-1$.
It can be transformed to the form
\begin{equation}
ds^2  = d\tau^2 - a^2(\tau)\frac{(d\zeta^1)^2+(d\zeta^2)^2+
(d\zeta^3)^2}{\left(1+ \frac{K}{4}\big((\zeta^1)^2+(\zeta^2)^2+
(\zeta^3)^2\big)\right)^2}, \label{g}
\end{equation}
that will be used below.

Let us consider the quaternionic algebra over
complex field ${\mathbb C}(1,\rm i)$ \cite{NS1}
\[
{\sf H}_{\cal C} = \{ \alpha + \beta i+ \gamma j + \delta k\; \mid
 \; \alpha , \beta,\gamma,\delta \in {\mathbb C} \}
\]
with multiplication operation defined by the property of bilinearity and
following rules for $i, \; j, \; k$:
\begin{eqnarray}
&&i^{2}=j^{2}=k^{2} =-1,\;\; jk=-kj = i, \nonumber\\
&&ki=-ik=j,\;\; ij=-ji= k.\nonumber
\end{eqnarray}
For the given quaternion $q=\alpha + \beta i+ \gamma j + \delta k$ its
quaternionic conjugate (denoted by $^{+}$) is given by
\[
q^+=\alpha - \beta i -\gamma j - \delta k,
\]
and one has $(q p)^+ = p^+ q^+$, $p,q\in{\sf H_{\cal C}}$.

Further we restrict ourselves to the set of quaternions
 ${\sf H}_{\mathbb R}$:
\begin{eqnarray}
{\sf H}_{\mathbb R} =&&\{\zeta=\zeta^0 + {\rm i}(\zeta^1 i+ \zeta^2 j+ \zeta^3
k):\; \;{\rm i}^2=-1,\;\rm i\in{\mathbb C}, \nonumber \\
&&\zeta^0,\zeta^1,\zeta^2,\zeta^3  \in {\mathbb R} \}  \nonumber
\end{eqnarray}
 with the norm $\|\zeta\|^2$ given by
\begin{equation}
\|\zeta\|^2=\zeta {\zeta}^+=(\zeta^0)^2 - (\zeta^1)^2-(\zeta^2)^2 -
 (\zeta^3)^2.
\end{equation}

Now let us consider one-parametric set of quaternions:
\begin{align*}
{\sf H}^{\mathcal F}_{\mathbb R} =\{x= \tau+\zeta^0 + a(\tau+\zeta^0)
\mbox{\boldmath$\zeta$}, \mbox{\boldmath$\zeta$}= {\rm i}(\zeta^1 i+ \zeta^2 j+
\zeta^3 k) \}
\end{align*}
where $\tau$ is a parameter. We define a binary operation $L_xy=x\ast y, \; x,y
\in {\sf H}^{\mathcal F}_{\mathbb R}$, depending on the parameter $\tau$, as
\begin{equation}
L_x y= \tau+ \eta^0 + \zeta^0 + a(\tau+\eta^0 +
\zeta^0)\big(\mbox{\boldmath$\zeta$}+\mbox{\boldmath$\eta$}\big)\Big/\Big(1 +
\frac{K}{4}\mbox{\boldmath$\zeta$}^+\mbox{\boldmath$\eta$} \Big), \label{QL}
\end{equation}
where $/$ denotes the right division. It is easy to check that $e=\tau$ is the
neutral element, $L_ex =x$. The inverse operation is given by
\begin{equation*}
L^{-1}_x y=\tau+ \eta^0 - \zeta^0 + a(\tau + \eta^0 - \zeta^0)
\big(\mbox{\boldmath$\eta$}-\mbox{\boldmath$\zeta$}\big)\Big/\Big(1 -
\frac{K}{4}\mbox{\boldmath$\zeta$}^+\mbox{\boldmath$\eta$} \Big),
\end{equation*}
and the associator is found to be
\begin{equation}
l_{(x,y)}z= \tau+\xi^0 + a(\tau+ \xi^0)
\Big(1 + \frac{K}{4}\mbox{\boldmath$\eta$}^+\mbox{\boldmath$\zeta$} \Big)\mbox{\boldmath$\xi$}
\Big/\Big(1 + \frac{K}{4}\mbox{\boldmath$\zeta$}^+\mbox{\boldmath$\eta$} \Big).
\end{equation}

Thus, the set of quaternions ${\sf H}^{\mathcal F}_{\mathbb R}$ with the binary
operation $\ast$ forms a loop, that will be denoted as $\rm Q{\sf H}^{\mathcal
F}_{\mathbb R}$. Notice, that $\rm Q{\sf H}^{\mathcal F}_{\mathbb R}$ admits a
natural geodiodular structure induced by the quaternionic algebra.

Employing Eq. (\ref{G1}), we define the left invariant diodular metric on $\rm
Q{\sf H}^{\mathcal F}_{\mathbb R}$ as follows:
\begin{equation}
\label{dg1} g^z(x,y)= \frac{1}{2}\Big(\big(L^{-1}_z x-e\big)\big(L^{-1}_z
y-e\big)^{+}+ \big(L^{-1}_z y-e\big)\big(L^{-1}_z x-e\big)^{+}\Big).
\end{equation}
In particular this yields
\begin{equation}
g^{z}(x,x) = \big(x^0-z^0\big)^2
-\frac{a^2(\tau+x^0-z^0)\big\|\mbox{\boldmath$\xi$}-
\mbox{\boldmath$\zeta$}\big\|^2}{\Big\|1
-\frac{K}{4}\mbox{\boldmath$\xi$}^{+}\mbox{\boldmath$\zeta$}\Big \|^2 } .
\label{dm}
\end{equation}

Let $\mbox{\boldmath$\xi$}= \mbox{\boldmath$\zeta$}+d\mbox{\boldmath$\zeta$}$
and $z^0= x^0 + d\tau$, then (\ref{dm}) leads to the FRW metric  of the Eq.
(\ref{g}):
\begin{equation*}
ds^2  = d\tau^2 - a^2(\tau)\frac{(d\zeta^1)^2+(d\zeta^2)^2+ (d\zeta^3)^2}{\left(1+
\frac{K}{4}((\zeta^1)^2+(\zeta^2)^2+ (\zeta^3)^2)\right)^2}.
\end{equation*}

\vspace{0.1in} {\em Discrete FRW spacetime}.- We start  with the description of
nonassociative 3-dimensional homogeneous spaces. Let us consider a finite set
$M = {\mathbb Z}^3_n = \{ {\bf p} =(p^i)\;|\; p^i \in {\mathbb Z}_n, i = 1,2,3,
n\in{\mathbb N}\}$ where ${\mathbb Z}_n = \{p=-n,\dots,n \}$ is the set of
integers. We define a partial loop  $\rm Q{\sf H}_{{\mathbb Z}^3_n}$ as a set
of quaternions
\begin{eqnarray}
{\sf H}_{{\mathbb Z}^3_n} =&&\{\zeta_{\bf p}={\rm i}\ell
(p^1 i+ p^2 j+ p^3 k): \; \ell = {\rm const},\nonumber \\
&&{\rm i}^2=-1,\;\rm i\in{\mathbb C};\;
{\bf p}  \in {\mathbb Z}^3_n\}
\end{eqnarray}
with the norm $\|\zeta_{\bf p}\|^2=\ell^2 p^i p_i = -\ell^2\big((p^1)^2
+(p^2)^2+(p^3)^2\big)$ and the binary operation ${\sf H}_{{\mathbb
Z}^3_n}\times {\sf H}_{{\mathbb Z}^3_n} \mapsto {\sf H}_{{\mathbb Z}^3_n} $
given  by
\begin{equation}
\zeta_{\bf p}\star\zeta_{\bf q}=\zeta_{\bf pq} =\big({\zeta_{\bf
p}+\zeta_{\bf q}}\big)\Big/\Big({1+ \frac{K}{4}\zeta^{+}_{\bf p}\zeta_{\bf q} }\Big),\; \;
\zeta_{\bf p},\zeta_{\bf q} \in {\sf H}_{{\mathbb Z}^3_n} \label{DLp}.
\end{equation}

We introduce a partial geodiodular finite space ${\cal M}_n $ at the neutral
element $e =0$ as follows:

$\bullet$ $\rm Q{\sf H}_{{\mathbb Z}^3_n}$ is the odule with the unary operation
multiplication induced by the quaternionic algebra over $\mathbb Z$

$\bullet$ ${\cal M}_n^{+}= {\sf H}_{{\mathbb Z}^3_n}$ is the osculating space
with the structure of vector space induced by the quaternionic algebra over
$\mathbb Z$.

For the duiodular 3-metric and associator we have
\begin{align}
&g^{\zeta_{\bf p}}(\zeta_{\bf q},\zeta_{\bf q}) =
\frac{\|\zeta_{\bf p}-\zeta_{\bf q}\|^2}
{\|1 - \frac{K}{4}\zeta^{+}_{\bf p}\zeta_{\bf q} \|^2 },\label{dm2}\\
&l_{(\zeta_{\bf p},\zeta_{\bf q})}\zeta_{\bf m}
=\Big(1+ \frac{K}{4}\zeta^{+}_{\bf q}\zeta_{\bf p} \Big )
\zeta_{\bf m}\Big/\Big(1+ \frac{K}{4}\zeta^{+}_{\bf p}\zeta_{\bf q}\Big  ).
\label{ass2}
\end{align}

Taking into account that for a symmetric space its elementary holonomy is
determined by the associator \cite{S5,S8}:
$$
h_{(\zeta,\eta)} \xi= l_{(\zeta,L^{-1}_\zeta \eta)}\xi,
$$
we find
\begin{eqnarray}
h_{(\zeta_{\bf p},\zeta_{\bf q})}\zeta_{\bf m}
=\Big(1- \frac{K}{4}\zeta_{\bf q}\zeta^{+}_{\bf p} \Big )
\zeta_{\bf m}\Big/\Big(1- \frac{K}{4}\zeta^{+}_{\bf q}\zeta_{\bf p}\Big  ).
\label{hol2}
\end{eqnarray}

The smooth 3-space could be regarded as the result of ``limit process of
triangulating'' when  $\ell n = \rm const$ and $\|\zeta_{\bf p}\|^2
\rightarrow\|\zeta\|^2$, while $\ell \rightarrow 0 ,\; n \rightarrow \infty$.

Let us consider ${\bf q = p + \mbox{\boldmath $\delta$}}, |\mbox{\boldmath
$\delta$} |\ll n$. Then we have $\zeta_{\bf q} = \zeta_{\bf p} + \Delta
\zeta_{\bf q}$ where $\Delta \zeta_{\bf q} = \ell(\delta^0 + {\rm i}(\delta^1
i+ \delta^2 j+ \delta ^3 k)) $. The diodular 3-metric (\ref{dm2}) takes the
form
\begin{equation}
g^{\zeta_{\bf p}}(\zeta_{\bf q},\zeta_{\bf q}) =
\frac{\|\Delta\zeta_{\bf q}\|^2}
{\|1 - \frac{K}{4}\zeta^{+}_{\bf p}\zeta_{\bf p} \|^2 }+ O(K\ell^2),
\label{dm4}
\end{equation}
and
\begin{eqnarray}
g^{\zeta_{\bf p}}(\zeta_{\bf
q},\zeta_{\bf q}) \rightarrow
ds^2  = -\frac{(d\zeta^1)^2+(d\zeta^2)^2+(d\zeta^3)^2}
{\Big(1 + \frac{K}{4}\big((\zeta^1)^2+(\zeta^2)^2+(\zeta^3)^2\big)\Big)^2},
\nonumber
\end{eqnarray}
while $\ell \rightarrow 0,\; n \rightarrow \infty$.

Let $\rm {\sf H}^{\mathcal F}_{{\mathbb Z}^3_n\times{\mathbb T}_n}$ be a set of
quaternions:
\begin{align*}
\rm {\sf H}^{\mathcal F}_{{\mathbb Z}^3_n\times{\mathbb T}_n}=\{x_{\mathbf p}=
\tau_n +  a(\tau_n )\mbox{\boldmath $\zeta$}_{\mathbf p},\;{\bf p}  \in
{\mathbb Z}^3_n ,\},
\end{align*}
where $\tau_n $ is a discrete parameter, and $ {\mathbb T}_n=\{\tau_n : \; n\in
\mathbb N\}$,  then a partial loop $\rm Q{\sf H}^{\mathcal F}_{{\mathbb
Z}^3_n\times{\mathbb T}_n} \simeq \rm Q{\sf H}_{{\mathbb Z}^3_n} \times{\mathbb
T}_n $ is defined as the loop with the operation $\ast$ given by
\begin{equation}\label{z1}
x_{\mathbf p}\ast x_{\mathbf q}=\tau_n + a(\tau_n )\mbox{\boldmath
$\zeta$}_{\mathbf p}\star\mbox{\boldmath $\zeta$}_{\mathbf q}
\end{equation}

We introduce FRW nonassociative discrete spacetime as
\begin{eqnarray}
{\cal M}^{\mathcal F}= \bigcup_{n\in \mathbb N}{\cal M}^{\mathcal F}_n  ,
\end{eqnarray}
where a partial geodiodular finite space ${\cal M}^{\mathcal F}_n $ consists of

$\bullet$ $\rm Q{\sf H}^{\mathcal F}_{{\mathbb Z}^3_n\times{\mathbb T}_n}$
being the odule with the unary operation multiplication induced by the
quaternionic algebra.

$\bullet$ ${\cal M}_n^{+}= {\sf H}_{{\mathbb Z}^3_n}\times {\mathbb T}_n$ being
the osculating space with the structure of vector space induced by the
quaternionic algebra over $\mathbb Z$.

$\bullet$ For the given $x_{\mathbf p}\in {\cal M}^{\mathcal F}_{n} $ and
$x'_{\mathbf q}\in {\cal M}^{\mathcal F}_{n'}, \; (n\leq n')$, the partial
operation is defined as
\begin{equation}\label{z1a}
x_{\mathbf p}\ast x'_{\mathbf q}=\tau_{n'} + a(\tau_{n'} )\mbox{\boldmath
$\zeta$}_{\mathbf p}\star\mbox{\boldmath $\zeta$}_{\mathbf q}.
\end{equation}

The left invariant metric  can be introduced on ${\cal M}^{\mathcal F}$ in the
same way as in Eq. (\ref{dg1}). In particular, we obtain
\begin{equation}
g^{x_{\bf p}}(x_{\bf p},x'_{\bf q}) =(\tau_{n'} - \tau_{n} )^2-
\frac{a^2(\tau_{n'})\|\zeta_{\bf p}-\zeta_{\bf q}\|^2} {\|1 -
\frac{K}{4}\zeta^{+}_{\bf p}\zeta_{\bf q} \|^2 }.\label{dm3}
\end{equation}
In the continuous limit this yields the metric of FRW spacetime given by Eq.
(\ref{g}), and
we see that $\tau_n$ can be interpreted as a `discrete time' of FRW model.\\

\noindent {\bf Concluding remarks}. We have applied the general theory of
nonassociative geometry for description of nonassociative continuous and
discrete FRW models.

The first feature of our model is  obvious absence of the the problem of
initial singularity. The other important feature is appearence of the natural
arrow of time as follows: Consider ${\cal M}^{\mathcal F}_n $ as a triangulated
topological space. For the given triangulation, we identify the elements
$x_{\bf p}\in {\cal M}^{\mathcal F}_n $  as the vertices of the simplices, $n$
being a number of vertices. Since the operation $\ast$ is a partial operation,
in general for some elements $x_{\mathbf p}, x'_{\mathbf q}\in {\cal
M}^{\mathcal F}_n$ one has
$$
x_{\mathbf p}\ast x_{\mathbf q}\mapsto x_{\mathbf p'}\not\in {\cal M}^{\mathcal
F}_{n}.
$$
Adding the point $ x_{\mathbf p'}$ to the set ${\cal M}^{\mathcal F}_{n} $ as a
new vertex, we obtain ${\cal M}^{\mathcal F}_{n+1}$. This leads to the partial
ordering
\begin{eqnarray}
\dots\prec{\cal M}^{\mathcal F}_n \prec {\cal M}^{\mathcal
F}_{n+1}\prec\dots\prec{\cal M}^{\mathcal F}_{n'}\dots
\end{eqnarray}
induced by the operation (\ref{z1}), and ${\cal M}^{\mathcal F}$ becomes a
finite, partially ordered set or {\em causal set} \cite{F,BMLS}. Thus the
causal ordering in FRW space is related to the refinement of the triangulation
of 3-dimensional space, and the evolution of FRW universe is interpreted in
terms of a sequence
\begin{eqnarray}
\dots\rightarrow{\cal M}^{\mathcal F}_n \rightarrow {\cal M}^{\mathcal
F}_{n+1}\rightarrow \dots\rightarrow{\cal M}^{\mathcal F}_{n'}\dots
\end{eqnarray}
This is close to the interpretation given in \cite{EM}, where the universe
evolution is treated in terms of a sequence of topology changes in the set of
$T_0$-discrete spaces realized as nerves of the canonical partitions of
3-dimensional compact manifolds.

\acknowledgements

We thank V.N. Efremov for discussions on the work cited above.

\end{document}